\documentclass[10pt, conference]{IEEEtran}

	\usepackage{graphicx}
	\usepackage{color}
	\usepackage{epstopdf}
	\usepackage{amsmath}
	\usepackage{float}
	\usepackage[font=small]{caption}
	\usepackage[square,sort,comma,numbers]{natbib}
	
\begin{document}
	\title{A Survey of Privacy Infrastructures and Their Vulnerabilities}
	\author{\IEEEauthorblockN{Yunfan Tian}
	\IEEEauthorblockA{College of Engineering\\
	Northeastern University, Boston, USA\\
	Email: tian.yun@husky.neu.edu\\}
	\and
	\IEEEauthorblockN{Xiang Zhang}
	\IEEEauthorblockA{College of Engineering\\
		Northeastern University, Boston, USA\\
		Email: zhang.xiang1@husky.neu.edu\\}
	}
	\maketitle
	\begin{abstract}
	Over the last two decades, the scale and complexity of Anonymous networks and its associated technologies grows exponentially as privacy has become a major concern of individuals. Also, some cyber attackers make use of privacy infrastructures including botnets and Tor to do illegal activities like drug, contraband or DDoS attack. However, anonymous networks are not perfect, there are some methods could exploit the vulnerabilities and track user information. In this paper, we analyze few of privacy infrastructures and their vulnerabilities.
	\end{abstract}
	
	\section{Introduction}
	Due to the rapid and development of Internet, privacy has become a major concern of people. Also, the open natural of Internet encouraged illegitimate activities such as stealing user personal information, attacking advertising companies, viewing sensitive content and so on. Hence, privacy infrastructure emerged as a solution, VPN and Tor are typical examples for such systems \cite{DBLP:journals/corr/Balasubramanian16l}. These systems are responsible for hiding the identities for communicating peers and content of Internet flows from eavesdroppers. From the adversaries perspective, an anonymous system should have:
	\begin{enumerate}
		\item \textbf{Unlink-ability}, the inability to link two or more items of interest. 
		\item  \textbf{Unobservability}, items of interest are indistinguishable from all other items.
	\end{enumerate}

	However, privacy infrastructure still have vulnerabilities. For example, Dan Egerstad, a Swedish hacker, set up a rogue Tor exit node and intercepted the data that came through it in 2007. He successfully intercepted the emails and mail passwords of government, embassy, NGO and corporate staffers. In this paper, we analyze VPN and Tor, and their possible vulnerabilities.
	
	We present a brief background and possible vulnerabilities for above topics.
	
	\section{VPN}
	\subsection{Background}
	Virtual Private Network (VPN) is one of the well-known privacy infrastructure that has been trending in the past few years. It is a private network server provides secured communication to the destination by encrypting the traffic between the user and server \cite{DBLP:journals/corr/PriyankaP15}. VPN is a convenient and efficient means of connecting to the private network. Even though VPN offers secured communication, mitigating the risk of man in the middle attack, users privacy is not guaranteed.
	
	\subsection{Vulnerabilities of VPN}
	\begin{enumerate}
		\item \textbf{VPN provider leak information}, There are many cases where VPN providers share users information for business profit. 
		\item  \textbf{Unreliability of VPN service}, many VPN providers only use Weak Security protocol and offer Basic Encryption System, which hackers can easily compromised. This put their users at risk.
		\item \textbf{Statistical Inference Attacks}, Statistical analysis of traffic patterns can compromise anonymity, i.e. the timing and size of packets.
	\end{enumerate}

	\subsection{VORACLE: an attack on VPN}
	This attack was presented by Ahamed Nafeez\cite{VORACLE}, a security researcher, in Black Hat and DEF CON security conferences this year. In this attack, adversary can recover HTTP traffic was sent through an encrypted VPN under certain condition.

	It is not a brand new attack. Actually, it is mixed of some existed cryptographic attacks, such as BREACH, TIME, and CRIME. In the previous years, attacker can recover data from a encrypted TLS connection, if the data was compressed before encrypted. Security researcher have deployed the fixes of this attack in 2012 and 2013. Similarly, this attack can still valid in some types of VPN. In those VPN, server or client will compress data before encrypted it. In VORACLE attack, target can be secrets, it can be cookie, pages, or some sensitive information. If VPN services was built based on OpenVPN, it can be target of VORACLE attack. The reason is that OpenVPN protocol uses default setting that compresses all data before encrypted it via TLS and after that, they will be send in VPN channel. Attacker will lure a user to a HTTP site, this site can be controlled by attacker or run attacker's malicious code, such as some malicious ads. In that way, attacker can steal user's secrets.

	VORACLE can be prevent by not using OpenVPN based services, or using some Chrome-based browsers to defend malicious codes. OpenVPN have modified their docs page but they didn't modify defaul setting. Now, OpenVPN group have added an alert in their project. They will discuss in the future can fix it after that.
	
	\section{Tor}
	\subsection{Background}
	Tor is a low latency anonymity overlay network that is known to be the most privacy tool. It's the largest, most well deployed anonymity preserving service on the Internet since publicly available at 2002. 
	
	The basic design is a mix network with improvements, all traffic is protected with anonymous layers of encryption. When packets are transmitted, a random path with three nodes are used to make sure no single node is know the complete transmission path. Furthermore, Tor provide additional services:
	\begin{enumerate}
		\item \textit{Perfect forward secrecy}, Tor connections are encrypted by TLS protocol, the client negotiates a new public key pair with each relay, original key pairs only used for signatures.
		\item  \textit{Introduces guards to improve source anonymity}, Tor selects 3 guard relays and uses them for 3 months, then select 3 new guard. Guard relays help prevent attackers from becoming the first relay.
		\item \textit{Takes bandwidth into account when selecting relays}, help to reduce the latency.
		\item  \textit{Introduces hidden services}, servers that are only accessible via the Tor overlay. Tor allows you to run a server and have people connect without disclosing the IP or DNS name, that helps hiding the destination of traffic.
	\end{enumerate}

	\subsection{Vulnerabilities of Tor}
	Connection between exit node and destination is not encrypted and hence exit nodes can observe the content of messages. Dan Egerstad, a Swedish hacker, set up a Tor exit node and successfully intercepted the data that came through it.
	
	Predecessor attack is also a popular attack on anonymous systems like Tor. Suppose there are N total relays in Tor network, M of which are controlled by an attacker, the attacker goal is to control the first and last relay. It's roughly $(M/N)^2$ chance overall, for a single circuit. However, client periodically builds new circuits, the attacker just looks for the most frequent predecessor. Since routers are uniformly chosen, it should be unlikely that you have the same predecessor for the same session unless that predecessor us actually the entry TOR router.
	
	Last, Roger Dingledline mentioned that Tor client has 25\% to 30\% probability to select 5 fastest tor routers as relays. In fact, 80\% traffic of Tor network is depend on 40 to 50 Tor routers. Providing 40 to 50 fast Tor routers is not difficult for any country's intelligence agency.
	
	\section{Side Channel Attack}
	\subsection{Background}
	Side channel attack is any attack based on obtaining information on computer system (memory), vulnerabilities of algorithm, such as timing information, power consumption etc. Side channel attack is hard to find but also hard to defend.
	
	\subsection{Side Channels in Multi-Tenant Environments}
	Professor Reiter from University of North Carolina was working on side channel attack for many years\cite{Reiter:2015:SCM:2808425.2808426}. In the previous years, he focused on attack on Timing channel and Storage channel.

	In some Cloud services, most of side channel attack is through cache attack, so we need to design some pattern or algorithm to defense these cache-based attack. 
	
	There are two example of cloud, IaaS Clouds and PaaS Clouds. Infrastructure as a service (IaaS)\cite{iaas} is one of computing infrastructure. IaaS is perfered in some certain scenarios, such as test and development, website hosting, storage, backup, and recovery, web apps, high-performance computing, big data analysis, etc. There are some advantages of IaaS:
	
	\begin{enumerate}
	\item \textit{Eliminates capital expense and reduces ongoing cost}
	\item \textit{Improves business continuity and disaster recovery}
	\item \textit{Innovate rapidly}
	\item \textit{Respond quicker to shifting business conditions}
	\item \textit{Focus on your core business}
	\item \textit{Increase stability, reliability, and supportability}
	\item \textit{Better security}
	\item \textit{Gets new apps to users faster}
	\end{enumerate}

	Platform as a service (PaaS)\cite{paas} is another kind of computing infrastructure. It often be used in development framework, analytics or business intelligence or additional servives. There are also some advantages of PaaS:
	
	\begin{enumerate}
		\item \textit{Cut coding time}
		\item \textit{Add development capabilities without adding staff}
		\item \textit{Develop for multiple platforms - including mobile - more easily}
		\item \textit{Use sophisticated tools affordably}
		\item \textit{Support geographically distributed development teams}
		\item \textit{Efficiently manage the application lifecycle}
	\end{enumerate}

	Here are three attacks can be work on these system: Prime-Probe Attacks, Flush-Reload Attacks\cite{7163050} and Password Reset Attack. The core of these attack is manipulating memory. Take Prime-Probe Attacks as an example, clear cache of target and add some pivot in cache. When the system want to read something in cache, attacker can calculate the target data address based on memory reading time.

	\section{Conclusion}
	In this paper, three kinds of privacy infrastructures, VPN, Tor and side channel, are discussed and the vulnerabilities of them are analyzed. Since Tor is the most important privacy infrastructure, it attracts many companies and organizations to investigate its vulnerability. Further research is recommended to how to exploit the vulnerabilities of Tor.

	\bibliographystyle{IEEEtran}
	\bibliography{wp_ref}
	\end{document}